\documentclass[a4paper,12pt,titlepage]{article}

\usepackage{graphics}
\usepackage[nooneline]{caption}
\usepackage{sistyle}
\usepackage{upgreek}

\usepackage{setspace}

\newlength\figurewidth
\textheight=240mm{}
\parindent=0mm{}
\parskip=\bigskipamount

\begin{document}

\title{Temperature-Dependent In-Situ LEIS Measurement of W Surface Enrichment by 250 eV D Sputtering of EUROFER}

\author{
\parbox{\textwidth}{
H. R. Koslowski$^{1,\P}$,
S. R. Bhattacharyya$^2$,
P. Hansen$^1$,
Ch. Linsmeier$^1$,\\
M. Rasi\'nski$^1$
P. Str\"om$^3$
\\
\\
\normalsize{}
$^1$Forschungszentrum J\"ulich GmbH, Institut f\"ur Energie- und Klimaforschung -- Plasmaphysik,
52428 J\"ulich, Germany\\
$^2$Surface Physics \& Materials Science Division, Saha Institute of Nuclear Physics, 1/AF Bidhan Nagar, Kolkata-700064, India\\
$^3$Department of Fusion Plasma Physics, School of Electrical Engineering and Computer Science,
KTH Royal Institute of Technology, SE-10044 Stockholm, Sweden\\
\\
$^{\P}$ Corresponding author, email: h.r.koslowski@fz-juelich.de
\\
ORCID IDs:\\
S. R. Bhattacharyya https://orcid.org/0000-0002-8210-1088\\
P. Hansen  https://orcid.org/0000-0003-0099-6293\\
H. R. Koslowski https://orcid.org/0000-0002-1571-6269\\
Ch. Linsmeier https://orcid.org/0000-0003-0404-7191\\
M. Rasi\'nski https://orcid.org/0000-0001-6277-4421\\
P. Str\"om https://orcid.org/0000-0001-9299-3262\\
}
}

\date{3.7.2018, accepted for publication in \em Nuclear Materials and Energy\em}

\maketitle

\doublespacing

\abstract{
Tungsten surface enrichment of EUROFER steel by \SI{250}{eV} deuterium sputtering is in-situ measured using low energy He$^{+}$ ion scattering spectroscopy.
The samples are irradiated at various temperatures between 300 K and 800 K with a deuterium atom flux of \SI{2e18}{m^{-2}s^{-1}} and maximum fluence up to \SI{1.1e23}{m^{-2}}.
The measurements at room temperature show a clear increase of tungsten surface density, but already at 520 K the observed enrichment is only half as large.
At a temperature of 800 K no tungsten surface enrichment is detectable.
The obtained data yield an upper limit of \SI{1.6}{eV} for the diffusion activation energy of tungsten in EUROFER.
}
\vfill
\vfill

\newpage
\section{Introduction}

Materials for the first wall of a fusion reactor are the topic of intense research.
Although Be is chosen as first wall material for the ITER tokamak, alternative options, involving other metals and alloys, are under consideration for a future power plant.
In this context the application of reduced activation ferritic-martensitic (RAFM) steels like e.g. EUROFER for application in recessed and less loaded areas is considered \cite{tosc01,bolt02,nora03}.
Analyses of manufactured EUROFER batches yield the following contents of the main alloying elements: \SI{8.9}{mass\%} Cr, \SI{1.1}{mass\%} W, \SI{0.44}{mass\%} Mn, \SI{0.19}{mass\%} V, \SI{0.14}{mass\%} Ta, \SI{0.12}{mass\%} C, and a few other elements with a total content below \SI{0.05}{mass\%} \cite{lind05}.
The envisaged operating temperature window when used as first wall material extends up to \SI{820}{K}.
Besides high heat loads, the wall material has to withstand impinging particle fluxes which makes materials composed from medium- and high-$Z$ elements more suitable due to their lower sputtering yields.
Particle fluxes onto the first wall of a future fusion power plant are estimated to be in the range \SI{2e19}{m^{-2}s{-1}} to \SI{2e21}{m^{-2}s{-1}} \cite{behr03}.
EUROFER steel contains \SI{0.4}{at\%} of W \cite{lind05} whose sputter yield at low energy (here: \SI{250}{eV}) D irradiation is more than two orders of magnitude smaller than for the major constituent Fe \cite{sugi16}, which is thought to result in the surface enrichment of W by preferential sputtering.

Roth et al. \cite{roth14} have conducted dedicated experiments to study W surface enrichment of EUROFER using the plasma device PISCES-A \cite{goeb84}.
The erosion yield for low energy deuterons at various temperatures is measured.
An observed decrease of sputtering yield is attributed to W surface enrichment by preferential sputtering and increased W surface concentrations are confirmed by Rutherford backscattering analysis of the exposed EUROFER samples, but the limited depth resolution give lower surface concentration than anticipated due to TRIDYN \cite{moel88} modelling.
However, a repeated exposure at elevated temperature (\SI{500}{\degC}) reveals that the erosion yield returns to similar values as expected for a pure Fe surface, indicating a strong influence of temperature.

Plasma exposure of EUROFER samples in the PSI-2 \cite{kret15} linear plasma device at a fluence of \SI{1e26}{eV} and an D ion energy of \SI{60}{eV} to \SI{70}{eV} result in a complex change of the surface morphology and the development of a grass-like topology \cite{rasi17}. Energy dispersive X-ray analysis shows W enrichment in these structures.

Recently, erosion studies of magnetron sputtered Fe-W layers, assuming them to be a model system for RAFM steels, have been performed using either an high current ion source \cite{sugi15} or the linear plasma device PSI-2 \cite{stef15}.
The investigation with the high current ion source proves that the reduction in sputtering yield is correlated to the W surface enrichment analysed by Rutherford back scattering, and the experiment in the linear plasma device confirms the W enrichment at low energy D erosion using glow discharge optical emission spectroscopy to analyse the samples after exposure.
The latter study finds strong surface morphology modifications resulting in a needle-like structure \cite{rasi16}.

Similar results are obtained for another RAFM steel, F82H, which has been exposed with a maximum D fluence up to \SI{1e26}{m^{-2}} in a linear plasma generator \cite{alim16}.
Again, the erosion yield decreases with increasing D atom fluence but this effect is counteracted by an enhanced sample temperature.
Post-irradiation x-ray photoelectron spectroscopy (XPS) analysis shows W enrichment in the near surface layer, and scanning tunnelling electron microscopy (SEM) measurements reveals changes of the surface morphology.

Medium energy ion scattering (MEIS) analysis performed on various Fe-W coated samples which are irradiated with \SI{200}{eV} deuterons at fluences up to \SI{1e24}{m^{-2}} confirms the W surface enrichment and shows that the enriched layer depth is up to \SI{10}{nm}, depending on the fluence \cite{stro16,stro17}. Recently, a similar series of experiments using EUROFER samples confirms the W surface enrichment seen in the PISCES-A experiment \cite{roth14}, and yield increasing enrichment even at temperatures of \SI{900}{K} and \SI{1050}{K} where, however, recrystallisation and cracking of the sample surface is observed \cite{stro18}.

The above mentioned experiments using Fe-W coatings as a proxy for RAFM steels are modelled using the Monte-Carlo code SDTrimSP, which has been augmented by a model for solid-state diffusion in order to account for the observed temperature dependence \cite{tous16}.
Modelling results confirm the experimentally observed trends, i.e. an increasing W surface density with deuterium atom fluence and the weakening of the enrichment with increasing sample temperature.
Furthermore, a particle flux dependence of the W surface enrichment is encountered.
Again, the simulations show that under irradiation conditions, which resemble those in laboratory and plasma experiments, the depth of the enriched zone amounts a few \SI{}{nm} and shows a strong influence of temperature driven diffusion.

Low energy ion scattering (LEIS) is an extremely surface sensitive method which permits quantitative elemental analysis of the surface composition \cite{nieh93,bron07}.
In many situations LEIS is predominantly sensitive to the outermost 1 to 2 monolayers and only in cases where reionisation plays a role ions scattered at few deeper layers are detected.
Even in those cases the decay length, i.e. the depth where the ion signal has decayed to $1/{\rm e}$ is about \SI{1}{nm} \cite{prim11}.
In the following we report on the investigation of W surface enrichment by \SI{250}{eV} D sputtering of EUROFER steel and in-situ analysis of the W surface density using LEIS.

\newpage
\section{Experiment}

\begin{figure}[htb]
\resizebox{\figurewidth}{!}{\includegraphics{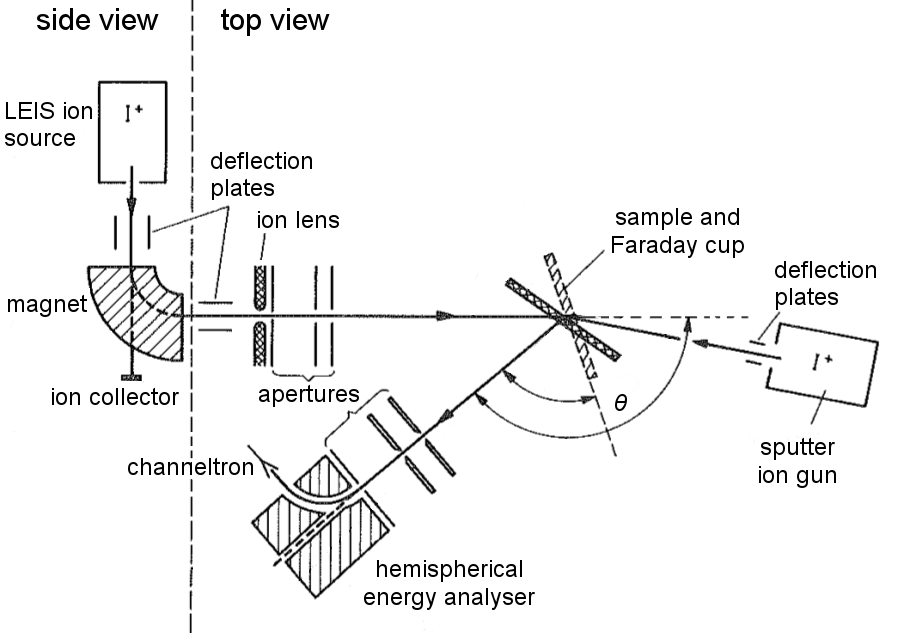}}
\caption{\label{fig-ALI} \doublespacing
Schematic drawing of the LEIS apparatus.}
\end{figure}

The experiments are carried out in an ultra-high vacuum chamber with a base pressure below \SI{5e-10}{mbar} measured with a Pfeiffer TPG 256 A MaxiGauge controller and a PBR 260 Bayard-Alpert hot cathode gauge.
A schematic drawing of the experimental setup is shown in figure \ref{fig-ALI}.
The apparatus has been originally developed and build at IPP Garching \cite{tagl85} and resumed operation after being moved to J\"ulich, recently.
The LEIS ion source is based on a Bayard-Alpert gauge and generates noble gas ions with a total ion current of several \SI{100}{nA}.
The ions are accelerated to an energy of typically \SI{1000}{eV} and focused with a first einzel lens mounted within the LEIS ion source before they pass through a sector magnet which selects the ion mass and charge state.
A second einzel lens focuses the probing ion beam onto the sample.
Two sets of deflection plates allow to steer the ion beam and correct for misalignments.
The ion current after mass/charge selection measured with a Faraday cup or on the sample is typically in the range \SI{1}{nA} to \SI{10}{nA}.
The ion beam diameter is about \SI{1}{mm}.
The EUROFER sample is mounted on a Prevac PTS 1000 RES/C-K sample holder which permits resistive heating up to $\sim$\SI{1200}{K}.
The sample holder is attached to a VG Scienta high precision 5-axis translator.
The ions which are reflected from the sample pass a hemispherical electrostatic energy analyser and are detected with a channeltron in single ion counting mode.
A multi-channel analyser has been programmed in Labview and uses a National Instruments PXI-8106 data acquisition system.
Sample cleaning is done with a Perkin-Elmer 04-161 sputter ion gun.
The diameter of the sputter ion beam is \SI{5}{mm} to \SI{6}{mm} and total ion currents are up to \SI{10}{\upmu A}.

The EUROFER samples have size \SI{10}{mm} $\times$ \SI{10}{mm} $\times$ \SI{1}{mm} and are initially polished to a mirror finish.
SEM images at magnifications up to 50000 show a smooth surface and grain sizes of a few \SI{}{\upmu m}.

Initial sample cleaning is performed by sputtering with 500 eV Kr$^{+}$ ions.
The total ion fluence used for cleaning, expressed as accumulated charge on the sample, is between \SI{2}{mC} and \SI{5}{mC}.
After this cleaning step the residual O contamination on the surface is very small and less than 1\% of the detected scattered He$^{+}$ ions originate from collisions with O.
It is well known that sputter cleaning of alloys may alter the surface composition of the sample already and yields surface densities of the various constituents which differ from their bulk densities.
In the simplified case of a binary alloy steady-state conditions are obtained when surface concentrations $c_{\rm s,x}$, bulk densities $c_{\rm b,x}$, and elemental sputtering yields $Y_{\rm x}$ obey the relation
\begin{equation}
 \frac{c_{\rm s,1}}{\rm c_{s,2}} = \frac{Y_{\rm 2}}{Y_{\rm 1}} \frac{c_{\rm b,1}}{c_{\rm b,2}}
\label{eq-patt}
\end{equation}
where the indices 1 and 2 label both elements of the alloy \cite{patt67}.
The sputtering yield of Fe for \SI{500}{eV} Kr bombardment is about 30\% larger than the yield for W \cite{yama96}, leading already after the sputter cleaning to a moderate W enrichment on the sample surface.
This small initial W enrichment on the surface does not cause any problems because the anticipated enrichment due to low energy D sputtering is much larger.

\begin{figure}[htb]
\resizebox{\figurewidth}{!}{\includegraphics{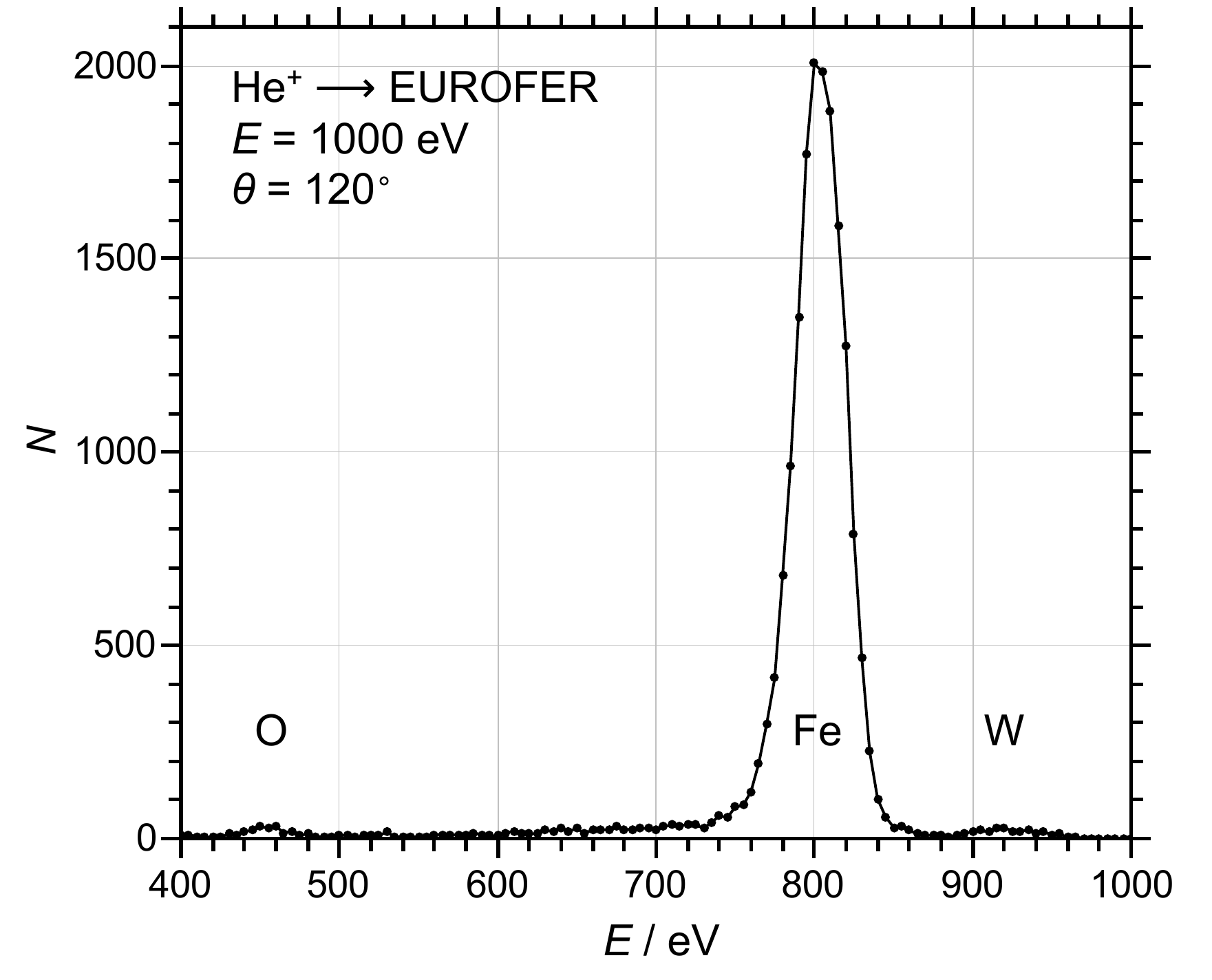}}
\caption{\label{fig-spec} \doublespacing
LEIS spectrum of 1 keV He$^{+}$ ions scattered on EUROFER.}
\end{figure}

A typical LEIS spectrum using singly charged He$^{+}$ ions accelerated to an energy of \SI{1000}{eV} measured at a scattering angle of 120$^{\circ}$ is shown in figure \ref{fig-spec}.
The LEIS spectrum shows a dominant peak resulting from scattering at Fe atoms, and well separated at higher energy a much smaller W peak.
A very small peak at \SI{450}{eV} is caused by the scattering from residual O atoms in the topmost layer of the sample.

Several series of D sputtering steps with increasing fluences are conducted at room temperature (RT), \SI{520}{K}, \SI{660}{K}, and \SI{800}{K}.
The D irradiation is done by introducing deuterium gas in the sputter ion gun, leading to an increase of the pressure in the vacuum vessel up to \SI{5e-6}{mbar} of D$_{2}$.
The sputter ion gun uses low energy electron impact ionisation of the working gas, thus producing mainly D$_{2}^{+}$ ions and up to 7\% of D$^{+}$ \cite{rapp65}.
The ions are accelerated with a voltage of \SI{500}{V}, thus yielding mainly \SI{250}{eV} deuterons impinging on the sample surface.
The sputter ion current is adjusted to get a flux density of \SI{2e18}{m^{-2}s^{-1}} on the sample.
The sputtering threshold energy for D atom bombardment of W is \SI{216}{eV} and the sputtering yield of Fe is at least 2 orders of magnitude larger than the sputtering yield of W for \SI{250}{eV} D projectiles \cite{sugi16}.
In-situ LEIS is conducted before and after each irradiation step.
After each D irradiation series at a certain temperature the sample surface is removed up to a sufficient depth by applying Kr$^{+}$ sputtering with a total ion current of \SI{1}{\upmu A} to \SI{2}{\upmu A} for about 1 hour, thus removing the enriched surface layer and resetting the sample.
Due to the slightly different sputtering yields for Fe and W \cite{yama96} the cleaning procedure leads already to a small W enrichment according to equation \ref{eq-patt} which results in a ratio of surface concentrations $c_{\rm W}/c_{\rm Fe}$ about 1.4 times larger than the respective ratio of bulk concentrations.

\begin{figure}[htb]
\resizebox{\figurewidth}{!}{\includegraphics{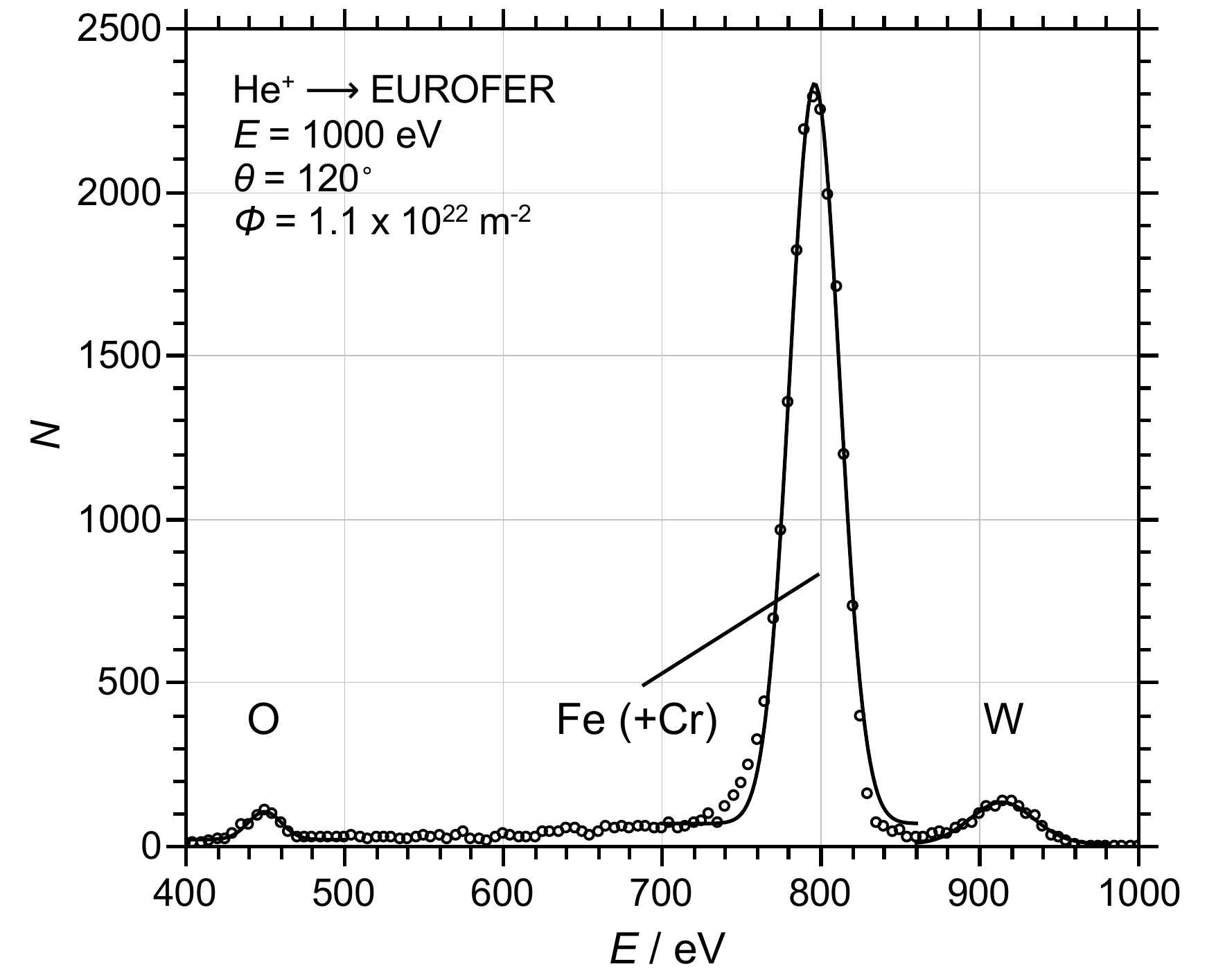}}
\caption{\label{fig-gauss} \doublespacing
LEIS spectrum measured after preferential sputtering using D with $E=\SI{250}{eV}$ and a fluence of $\mathit{\Phi} = \SI{1.1e22}{m^{-2}}$. Gaussian profiles are fitted to the O, Fe, and W peaks.}
\end{figure}

The three peaks in the LEIS spectra originating from scattering at O, Fe, and W, are fitted with Gaussian profiles.
An example spectrum with fitted peaks is shown in figure \ref{fig-gauss}.
Due to the composition of EUROFER, there arises one complication which cannot be accounted for easily.
EUROFER steel contains \SI{9}{mass\%} of Cr and the mass of a Cr atom is quite close to the mass of an Fe atom.
The energy resolution of the LEIS setup (energy broadening of the incident ion beam folded with the energy resolution of the electrostatic energy analyser) is not good enough to separate He$^{+}$ scattered at Fe and Cr.
The Cr peak is at the left wing of the Fe peak, resulting in a slight asymmetry as can be seen in figure \ref{fig-gauss}.
However, the error introduced by this superposition of scattering peaks is acceptable and, as can be seen in the figure, the area under the fitted peak is still a good representation for the number of scattered ions from Fe and Cr together.
A similar issue is present for the measured scattering events from W atoms, where the small Ta content gives an additional, not discernible contribution to the measured amplitude of the W peak.

The relative surface fraction of W is determined as the ratio between counts in the W peak (defined by the area under the peak) divided by the sum of counts in all measured peaks.
This definition is correct in case elemental sensitivities of all components are equal.
When elemental sensitivities are different, some systematic error is introduced.
However, the main constituent of EUROFER is Fe, the O contamination of the surface is small, and the surface fraction (i.e. the areal coverage) of W stays always much smaller than the surface fraction of Fe.
Under these constraints the count ratio mentioned above is a valid approximation for an uncalibrated ratio of surface fractions.
An absolute calibration is derived from a comparison of He$^{+}$ scattering spectra on pure Fe and pure W.

Surface observations are performed using a field emission scanning electron microscope (FE-SEM) Zeiss Crossbeam 540 equipped with energy-dispersive X-ray spectroscopy (EDX).
SEM investigations are carried out using a \SI{3}{kV} electron beam with current of \SI{172}{pA}.
The low electron energy enables measurements which are very sensitive to the surface.

\newpage
\section{Results and discussion}

\begin{figure}[htb]
\resizebox{\figurewidth}{!}{\includegraphics{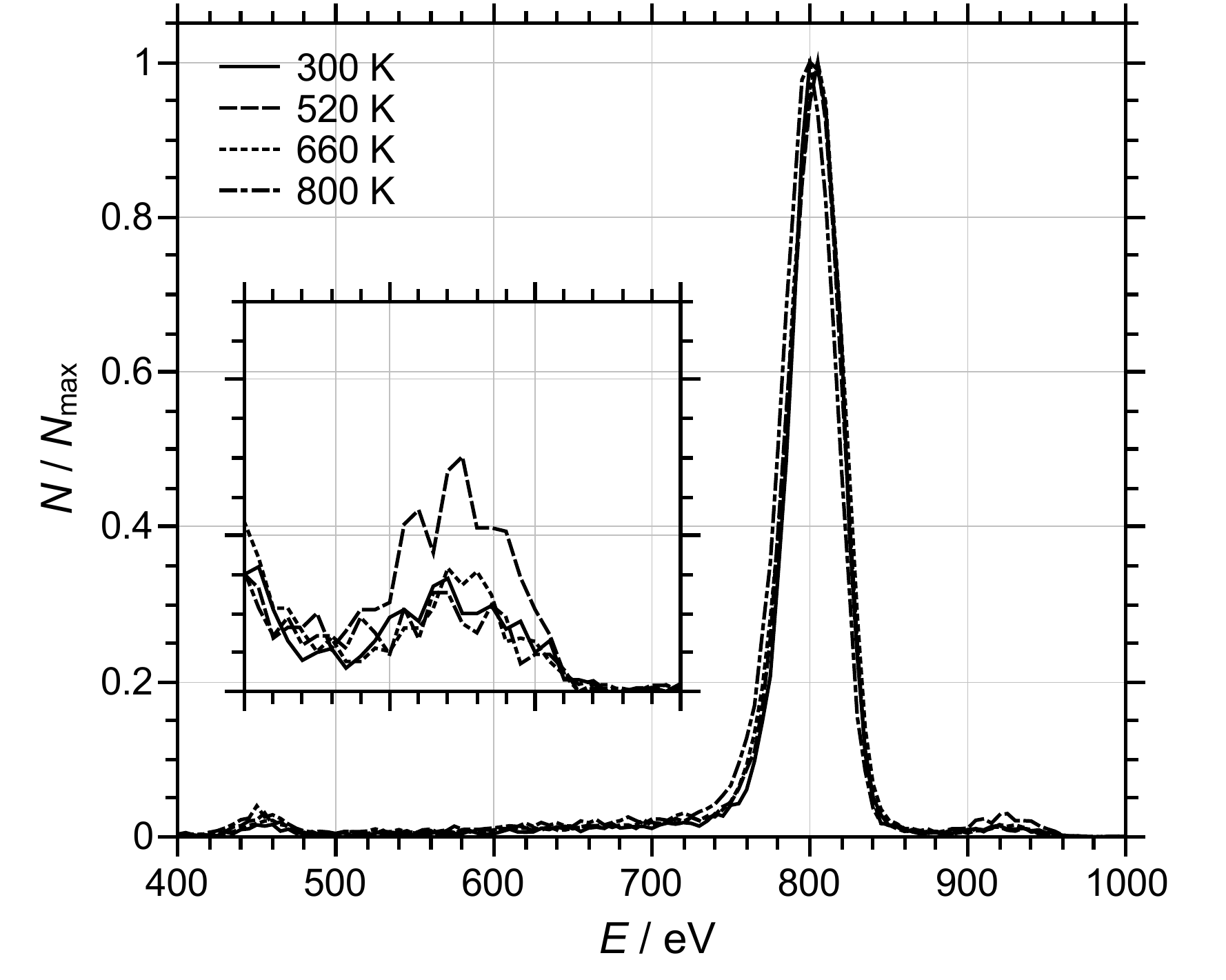}}
\caption{\label{fig-seg} \doublespacing
Normalised He$^{+}$ scattering spectra at \SI{300}{K}, \SI{520}{K}, \SI{660}{K}, and \SI{800}{K} showing that no surface segregation of W occurs at elevated temperatures.}
\end{figure}

One fundamental question to answer with respect to a measurement of surface enrichment of a specific alloy at various temperatures is, whether surface segregation occurs and might superimpose or interfere with the enrichment caused by sputtering.
The LEIS spectra after cleaning (and resetting the surface between irradiation sequences at different temperatures) are shown in figure \ref{fig-seg}.
The spectra are normalised to the Fe peak at \SI{800}{eV}.
There is always a small O peak, which tends to increase very little at elevated temperatures.
This behaviour may be due to the fact that the time between sample cleaning and LEIS measurements increases when the sample has to be heated up to the target temperature, thus allowing for a longer time O deposition from the residual gas.
The W peaks at \SI{920}{eV} are very small and do not show a systematic increase with temperature.
However, the measurement at \SI{520}{K} shows a larger W peak which is believed to be caused by insufficient sample cleaning before the heating.
It is important to note that this larger W fraction on the sample surface as the starting point for the irradiation sequence at \SI{520}{K} does not introduce any difficulty for the subsequent irradiation measurements because the W surface fractions obtained after each irradiation step are larger.
The absence of an increase of the W peak shows that for the temperatures used in these experiments surface segregation does not play a role.
This conclusion is further supported by density-functional theory (DFT) calculations of the segregation energies in transition-metal alloys, which for the segregation of W in Fe yield a small but positive surface segregation energy, thus even favouring a moderate anti-segregation \cite{ruba99}.
However, this result of the DFT calculations can only be used as a weak confirmation since EUROFER is not a binary alloy but has multiple components and may behave differently.
The spectra show small shifts of the centre position of the Fe peaks which is likely caused by small changes in the ion beam parameters, especially optimisation of the extraction voltage in order to get maximum ion current does lead to a change in beam energy.

\begin{figure}[htb]
\resizebox{\figurewidth}{!}{\includegraphics{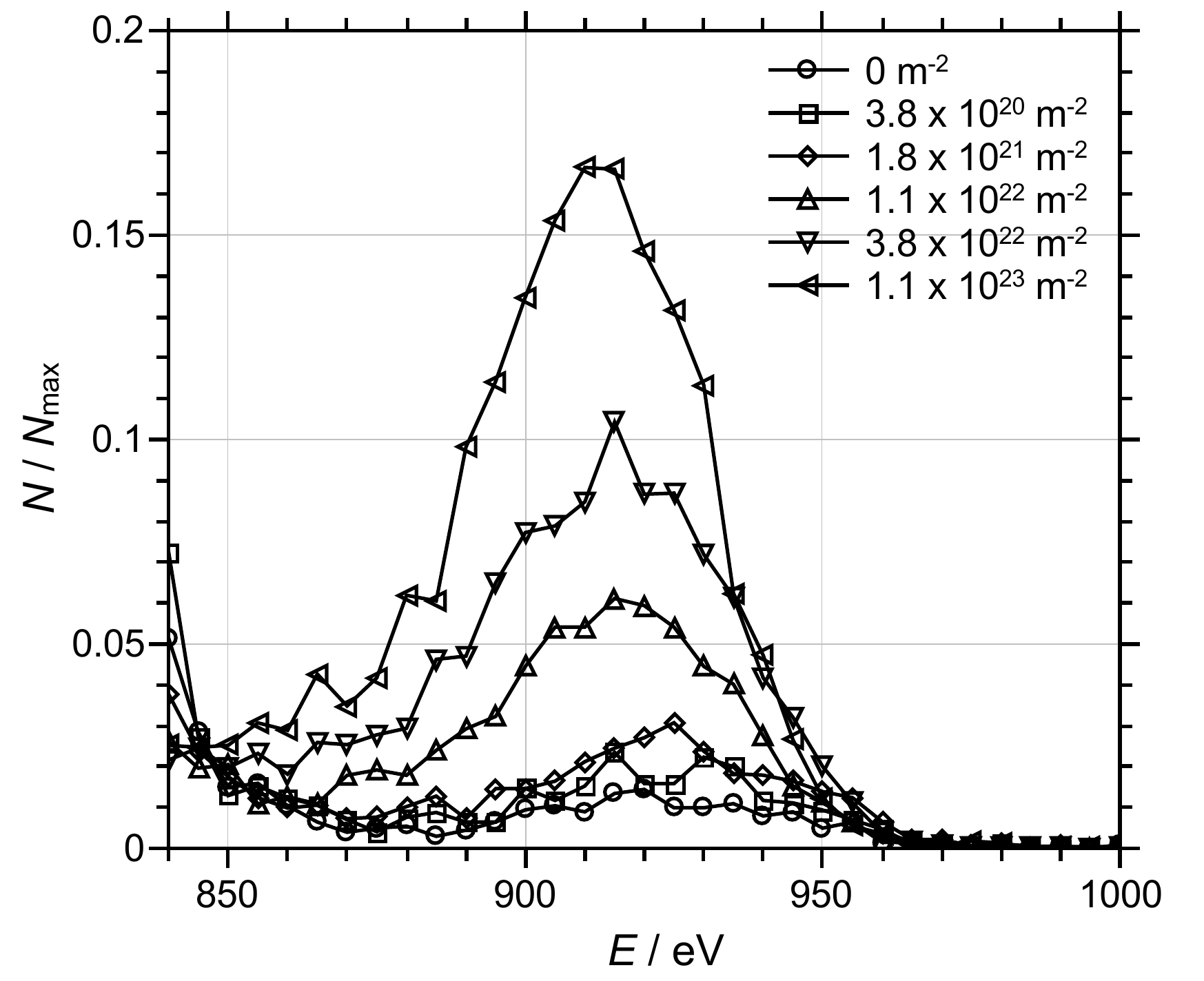}}
\caption{\label{fig-norm} \doublespacing
High energy range of several LEIS spectra showing the W peak in a series of consecutive measurements taken with increasing D fluence. The sample is at room temperature. All spectra are normalised to the height of the Fe peak.}
\end{figure}

Figure \ref{fig-norm} shows the W peak of six LEIS spectra for the irradiation sequence at RT.
The spectrum labelled with a fluence of \SI{0}{m^{-2}} is measured after sputter cleaning prior to the D bombardment, four spectra are measured at various, increasing fluence values, and the sixth spectrum is measured at the maximum D atom fluence of \SI{1.1e23}{m^{-2}}.
The spectra are normalised to the height of the Fe peak and for better visibility only the high energy part of the spectrum with the W peak is shown.
The W peaks exhibit a clear correlation with the increase of the D atom fluence.
The relative peak height of the W peak is about 1\% of the Fe peak for the cleaned sample and reaches a value of 16\% at the maximum applied D fluence.

\begin{figure}[htb]
\resizebox{\figurewidth}{!}{\includegraphics{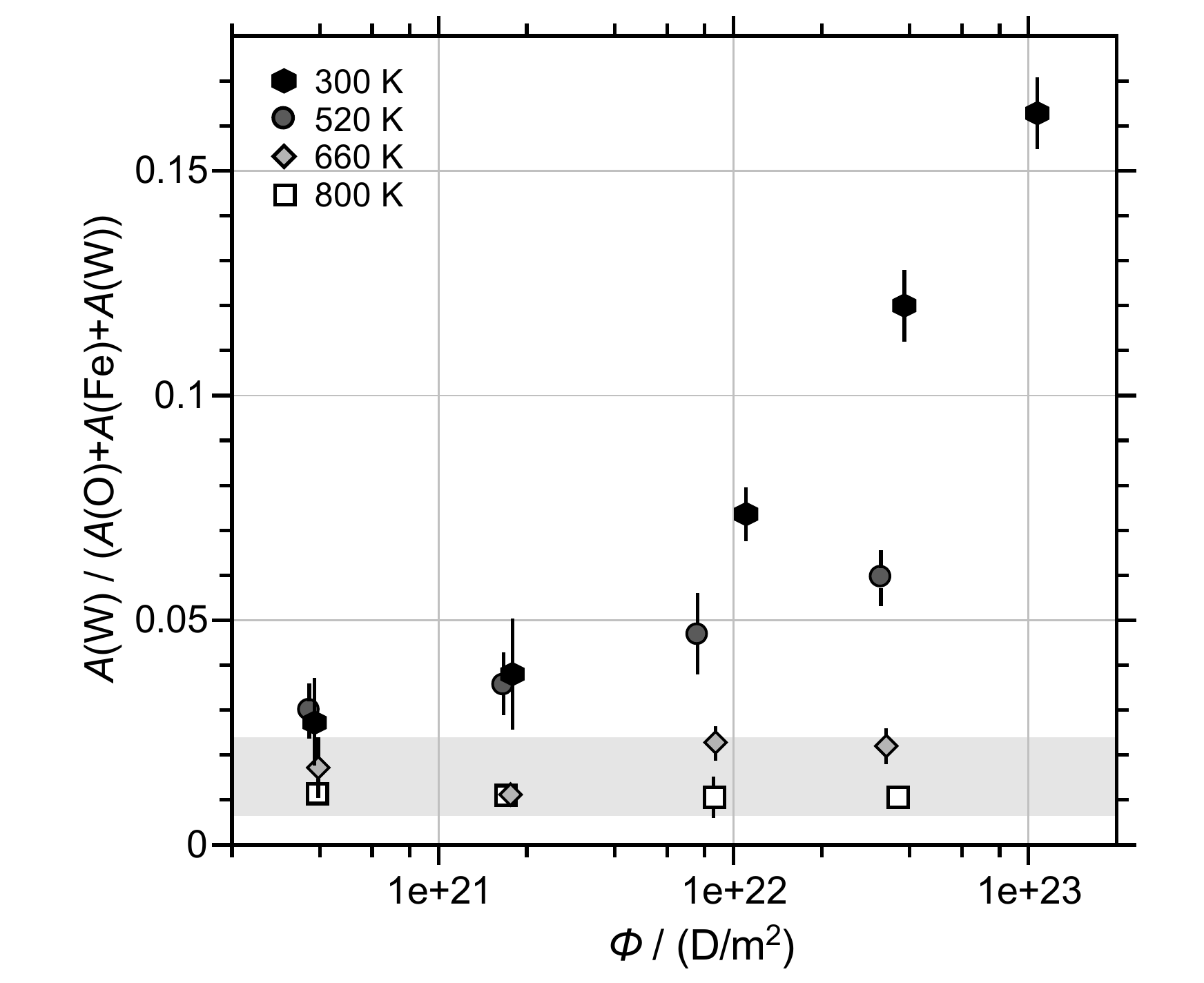}}
\caption{\label{fig-enrich} \doublespacing
Relative surface density of W vs fluence of 250 eV D atoms at various temperatures.}
\end{figure}

Similar irradiation sequences are repeated for elevated sample temperatures.
The measured W enrichments on the surface are summarised in figure \ref{fig-enrich}.
The shaded area marks the range of surface densities measured for the various sequences just after sample cleaning and heating to the desired temperature.
The upper limit is distorted due to the insufficient sample cleaning at \SI{520}{K} and the measured relative densities are about $(0.01\pm 0.005)$.
At RT W shows the strongest surface enrichment up to a relative value of $0.163\pm 0.008$.
No sign for saturation of the W surface density at the maximum D fluence of \SI{1.1e23}{m^{-2}} can be seen.
This finding is in good agreement with investigations of D sputtering of Fe/W coatings where an increase of the W surface density up to 24\% with a fluence of \SI{1e24}{m^{-2}} has been detected \cite{stro17}.
With increasing temperature the measured surface enrichment at each irradiation step decreases.
Already at a sample temperature of \SI{520}{K} the surface enrichment is only half as large as at RT, and at \SI{800}{K} the D$_{2}^{+}$ bombardment does not lead to any significant increase of the W surface density.
The experimental results prove that it is possible to demonstrate in-situ the W enrichment of EUROFER steel by preferential sputtering using D atoms with an energy close to the sputter threshold.

The LEIS data presented above give relative surface densities by not differentiating between differential scattering cross sections and neutralisation probabilities.
The count rate of particles scattered from atoms of element i on the surface, ${\dot N_{i}}$ is given by
\begin{equation}
{\dot N_{i}} = \frac{{\rm d}\sigma_{i}}{{\rm d}\Omega} {\dot N_{0}} c_{i} P^{+}_{i} T \Delta\Omega{}
\label{eq-scat}
\end{equation}
where ${{\rm d}\sigma_{i} / {\rm d}\Omega}$ is the differential cross section, ${\dot N_{0}}$ the rate of incident ions on the surface, $c_{i}$ the surface density of element i, $P^{+}_{i}$ is the ion fraction, i.e. the probability that the scattered ion is not neutralised, $T$ is the transmission and sensitivity factor of the analyser, and $\Delta\Omega$ the detected solid angle \cite{nieh93}.
From equation \ref{eq-scat} it follows that only the product of differential cross section and ion fraction depends on the individual element i, which permits to define the elemental sensitivity according to \cite{bron07}
\begin{equation}
\eta_{i} = P^{+}_{i} \frac{{\rm d}\sigma_{i}}{{\rm d}\Omega}.
\end{equation}
When resonant neutralisation or Auger neutralisation prevail and collision induced processes are neglected, the ion yield can be expressed as the product of the scattering cross section and an ion survival probability which is given by $P^{+} = \exp(-v_{c}/v_{\perp})$, where $v_{c}$ is a characteristic velocity and $v_{\perp}$ is the effective value of the velocity of the projectile perpendicular to the surface, taking into account incoming and outgoing trajectories.
However, when more than one process contribute, the fraction of scattered ions cannot anymore be described by a survival probability, but is given by
$P^{+} = P^{+}_{\rm in} (1 - P_{\rm CIN}) P^{+}_{\rm out} + (1 - P^{+}_{\rm in}) P_{\rm CIR} P^{+}_{\rm out}$, where $P^{+}_{\rm in}$ and $P^{+}_{\rm out}$ are the survival probabilities on the ingoing and outgoing trajectories, and $P_{\rm CIN}$ and $P_{\rm CIR}$ are the probabilities for collision induced neutralisation and reionisation, respectively \cite{bron07}.

In order to get absolute surface fractions for W, the ion yields for scattering from W atoms and Fe atoms need to be determined.
For that purpose a special sample is prepared.
Half of a pure Fe sample is covered before magnetron sputtering a thin layer of W onto the sample.
The calibration sample is mounted on the manipulator in a way, that by vertical translation the probing ion beam is scattered either from a pure Fe surface or from a pure W surface.
In that way all other quantities in equation \ref{eq-scat} are kept constant and the number of detected scattered particles in both spectra is according to the elemental sensitivities for both elements.
Small O peaks on both calibration surfaces are taken into account by normalisation of the Fe or W count rates to the total count rates for the respective surface.
As already discussed for the evaluation of the W surface fraction of EUROFER, the assumption of equal elemental sensitivities is only correct in first order and may introduce an error.
However, the W scattering spectrum shows the surface peak and an extended wing at the low energy side which indicates that a fraction of the scattered ions are reionised \cite{prim11}.
The influence of this feature on the elemental sensitivity factor has been minimised by fitting only the high energy half of the peak and avoiding an unreasonably large half width.
This approach is justified by neutral scattering spectroscopy measurements of \SI{1}{keV} He bombardment of W which has shown that the peak of scattered and ionised He$^{+}$ is displaced by \SI{14}{eV} to the low energy side \cite{thom86}.
It is important to note that this measurement includes ionised neutrals scattered at all layers, especially the dominant fraction with lowest energy loss which has been scattered on the surface.
Reionised particles which are scattered at the first (or second) monolayer and feature low energy losses, i.e. superimpose the direct scattering peak and are counted by the peak fitting procedure, are of no concern because they detect surface atoms.
The energies of particles scattered at deeper layers (which is important in this context, because these particles could distort the surface density determination) are shifted even more toward the low energy side \cite{drax03} and are more effectively excluded by the chosen fitting procedure.
However, an estimation of the magnitude of the error which is introduced by these sub-surface scattering events cannot be determined from the calibration spectrum and one has to rely on the facts that (i) the energy of those particles moves progressively away from the surface peak the deeper the scattering takes place \cite{drax03}, i.e. these ions do not contribute to the measured surface peak, and (ii) their number decrease rapidly with depth and limits the detected signal to a few monolayers with the dominant signal contribution originating from the topmost monolayer \cite{prim11}.
The ratio of the number of counts in the fitted W peak and the Fe spectrum is $4/3$.
Due to the unavoidable contribution of some ions scattered from deeper layers the determined sensitivity factor constitutes an upper limit.
The maximum of the relative W surface fraction for the room temperature measurement at highest fluence shown in figure \ref{fig-enrich} is \SI{16.3}{\%}.
Correcting for the slightly higher sensitivity for detection of W and considering the presence of reionised particles yields a surface coverage of up to \SI{12}{\%}.

The relative quantification of the scattered ion signals of the various constituents is complicated by the fact that He$^{+}$ scattering from Fe and from W at the chosen scattering energy takes place in different regimes due to their different thresholds for reionisation.
The scattering on Fe is dominated by Auger neutralisation, whereas scattering from W is in the reionisation regime \cite{bron07}.
Furthermore, due to the small O content on the surface the presence of a matrix effect cannot be excluded.
The Fe peak (see figure \ref{fig-spec}) shows almost no contribution from reionised projectiles (which are expected to have lower energies, i.e. are located in the left wing of the peak), but the W peaks in figure \ref{fig-norm} show a pronounced left wing (see e.g. the data for maximum sputter fluence, left pointing triangles).
The normalised scattering intensity at \SI{850}{eV} (about \SI{60}{eV} below the direct scattering peak) is 0.025 and does still contain a contribution of ~0.01 from the right wing of the Fe scattering peak.
The remaining amplitude of 0.015 is less than 10\% of the W peak amplitude.
The same determination of the magnitude of reionisation for the pure W spectrum measured with the calibration sample yields a value of about 1/3, i.e. a much larger contribution.
Obviously, the influence of reionisation in EUROFER, where W is a minor constituent and the achieved surface enrichment is not very large, is much smaller than in the pure element.
This questions the conclusion drawn in \cite{goeb15} where the comparison with data from a pure reference sample has been assessed to be sufficient for quantification.
From the similar shape of the W peaks in figure \ref{fig-norm} one can conclude that the contribution of reionised scattered ions for all irradiation steps is similar and the relative increase of surface concentrations is still correct.

A matrix effect due to the presence of O on the surface, as for example documented in \cite{kuer13} for a NiO surface or measured in \cite{nieh75} for WO, has not been detected in the present experiments.
The O exposure at a base pressure of less than \SI{5E-10}{mbar} amounts to \SI{8}{L} for the measurement at RT with maximum fluence (which needs about 30 hours), and is \SI{2}{L} to \SI{3}{L} for the measurements with lower fluences at elevated temperatures.
During most of the time the sample is under sputter bombardment, i.e. the adsorbed oxygen is removed immediately.
Furthermore, the heated samples do not show an increase of the O peak at all but at \SI{520}{K} the W peak still increases.
No decrease of the W signal with increasing measurement time (i.e. increasing O exposure) is seen and the variation of count rate for the Fe peak (which is within a factor of 2 for all measurements) is compatible with variations of the probing beam intensity between different measurements.

\begin{figure}[htb]
\resizebox{\textwidth}{!}{\includegraphics{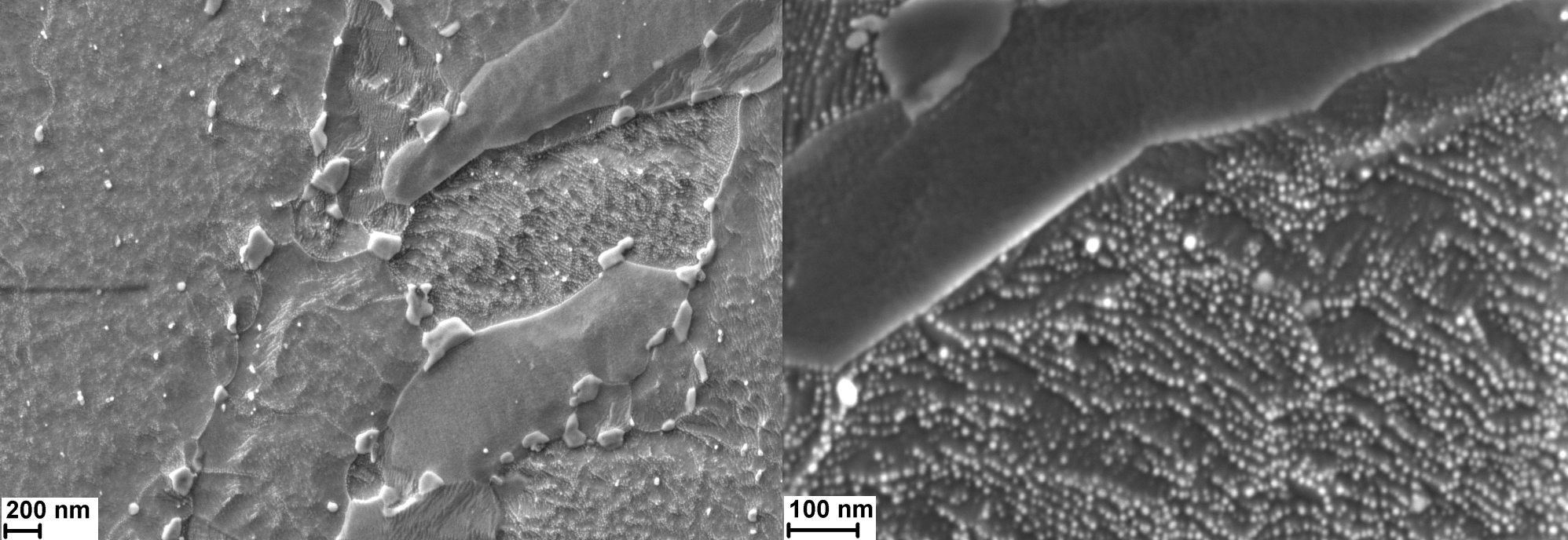}}
\caption{\label{fig-sem} \doublespacing
SEM images of EUROFER sample after \SI{250}{eV} D bombardment with a fluence of $\mathit{\Phi} = \SI{1.1e23}{m^{-2}}$ with two magnifications.}
\end{figure}

Two scanning electron microscopy images of the EUROFER surface after exposure at the maximum fluence are shown in figure \ref{fig-sem}.
One can notice different sputtering behaviour depending on the grain orientation.
Two types of sputtered grain morphology can be observed.
Part of the EUROFER surface is very smooth and flat, whereas other grains exhibit formation of nanoparticles with size below \SI{5}{nm}.
Based on the EDX analysis one can attribute these particles to oxide formation.

The sample at maximum W enrichment is additionally analysed with energy dispersive X-ray spectroscopy (EDX), X-ray photoelectron spectroscopy (XPS), and time-of-flight medium energy ion scattering (TOF-MEIS) spectroscopy.
All three methods facilitate to determine the elemental composition, but have larger information depths than LEIS.

The EDX analysis utilises an Oxford XMax-80 detector and the Oxford AZTec software package which applies the Pouchou and Pichoir matrix correction method \cite{pouc87}.
EDX gives values between $(\num{1.0} \pm \num{0.1})\,\SI{}{at\%}$ and $(\num{1.1} \pm \num{0.1})\,\SI{}{at\%}$ depending on the selected spot on the sample.
The values are almost three times larger than the W bulk concentration of \SI{0.4}{at\%} in EUROFER.
However, the detected level of enrichment compared to the LEIS measurements is still quite small, resulting from the fact that the EDX measurement averages over a \SI{50}{nm} to \SI{60}{nm} thick layer.
Assuming that the actual thickness of the enriched layer is much smaller, as indicated e.g. in the SDTrimSP modelling \cite{tous16}, the EDX measurement does not conflict with the ion scattering result.

The XPS measurements of the elemental composition of the surface of the same sample are summarised in table \ref{tab-xps}.
Since the enriched layer thickness is very small and sputter cleaning is known to strongly change the surface composition when concentrations and sputter yields do not obey the equilibrium relation given by equation \ref{eq-patt}, no sample cleaning has been done before the measurements.
It is therefore no surprise that the XPS data show dominant C and O contaminations due to the storage and sample transport in air as is shown in the first row of table \ref{tab-xps}.
EUROFER steel contains about \SI{0.5}{at\%} C and less than \SI{0.005}{at\%} O.
When O is excluded from the elemental analysis, the W concentration rises to \SI{5.8}{at\%}.
When C and O are excluded, the W concentration becomes \SI{9.3}{at\%}.
The actual W concentration on the surface is in the range given by the latter two values. Compared with the LEIS measurement both values are smaller due to the larger information depth of the XPS measurement which is in the range \SI{5}{nm} to \SI{10}{nm}.

\begin{figure}[htb]
\resizebox{\figurewidth}{!}{\includegraphics{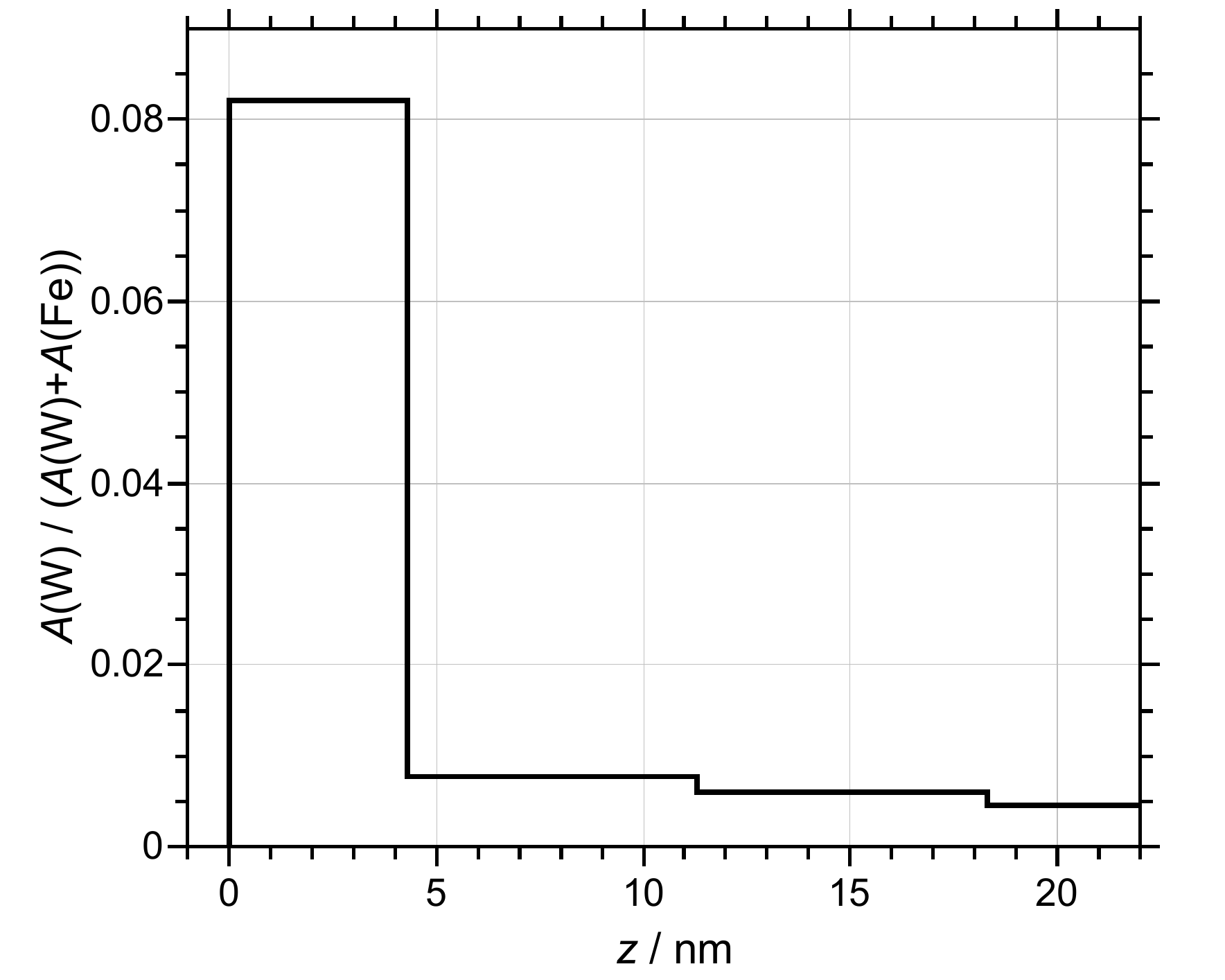}}
\caption{\label{fig-meis} \doublespacing
MEIS analysis of EUROFER sample after \SI{250}{eV} D bombardment with a fluence of $\mathit{\Phi} = \SI{1.1e23}{m^{-2}}$}
\end{figure}

The result of the TOF-MEIS measurement is shown in figure \ref{fig-meis}.
The applied MEIS setup is described in detail in \cite{linn12}.
A He$^{+}$ beam with \SI{60}{keV} is used for the measurement and scattered ions are detected under an angle of \SI{155}{\arcdeg}.
Three individual measurements are taken and averaged to obtain a spectrum with better statistics.
The resulting spectrum is compared to simulations using the TRBS code \cite{bier91}.
A reasonable estimate for the depth resolution of the MEIS measurement is \SI{2}{nm} based on the energy resolution of \SI{1}{keV} of the time-of-flight system.
The quantitative analysis of the TOF-MEIS spectra follows the procedure outlined in \cite{stro18}.
The thicknesses of the various layers and the concentrations of O, Fe, and W are adapted until satisfactory agreement with the measured spectrum is obtained.
Similar to the LEIS measurements, the measured continuum for scattering events resulting from Fe and Cr (and V, Mn) are summed up and attributed to Fe, scattering events resulting from W and Ta are both attributed to the W signal.

There are two free parameters required when setting up the TRBS simulation. These are the atomic fractions of W and O, the atomic fraction of Fe follows from the condition that all fractions should add up to one.
The measured quantities are the Fe and W signal amplitudes in the spectrum, therefore both free parameters can be clearly determined.
This is true for every depth, i.e. every corresponding energy interval.
The relationship between energy spectrum and depth profile can be non-ambiguously obtained.
The signal normalisation is done in the energy range \SI{20}{keV} to \SI{30}{keV} at sample depths beneath \SI{20}{nm} assuming an unmodified composition.
The above reasoning is weakened if other elements than Fe (+ Cr, V, Mn), W (+Ta), and O are present.
However, even in this case the $A(\rm Fe)/(A(\rm Fe)+A(\rm W))$ ratio is accurate when the O fraction is interpreted to represent "everything else", e.g. C.
A small error is introduced by this assumption because the stopping power of other elements is different.

The result shows a $A_{\rm W}/(A_{\rm Fe}+A_{\rm W})$ ratio of roughly 8\% and a thickness of the enriched layer of about \SI{4}{nm}. As with the XPS analysis, no sputter cleaning of the sample is done.

The W surface concentration measured with LEIS is larger than the values determined by any of the other methods, proving the extreme surface sensitivity of LEIS.

The results from LEIS and the other methods show that the measured value of W surface enrichment depends on the information depth of the applied analysis method.
The larger the information depth, the smaller is the obtained surface density, showing that the W enrichment happens in a thin layer.
However, the extreme surface sensitivity of LEIS raises the question what the outermost atomic layer in reality is, especially in view of the fact that crystal orientation can influence the scattering signal \cite{prim08} and the surface exhibits some strong features of morphology changes after the irradiation sequences (as can be seen in the SEM images).
For simplicity we adopt the point of view that incident ions are scattered on the same surface layers as would be particles from the plasma under real exposure conditions.
In that respect the ion beam experiment is relevant and mimics the intended application.

\begin{figure}[htb]
\resizebox{\figurewidth}{!}{\includegraphics{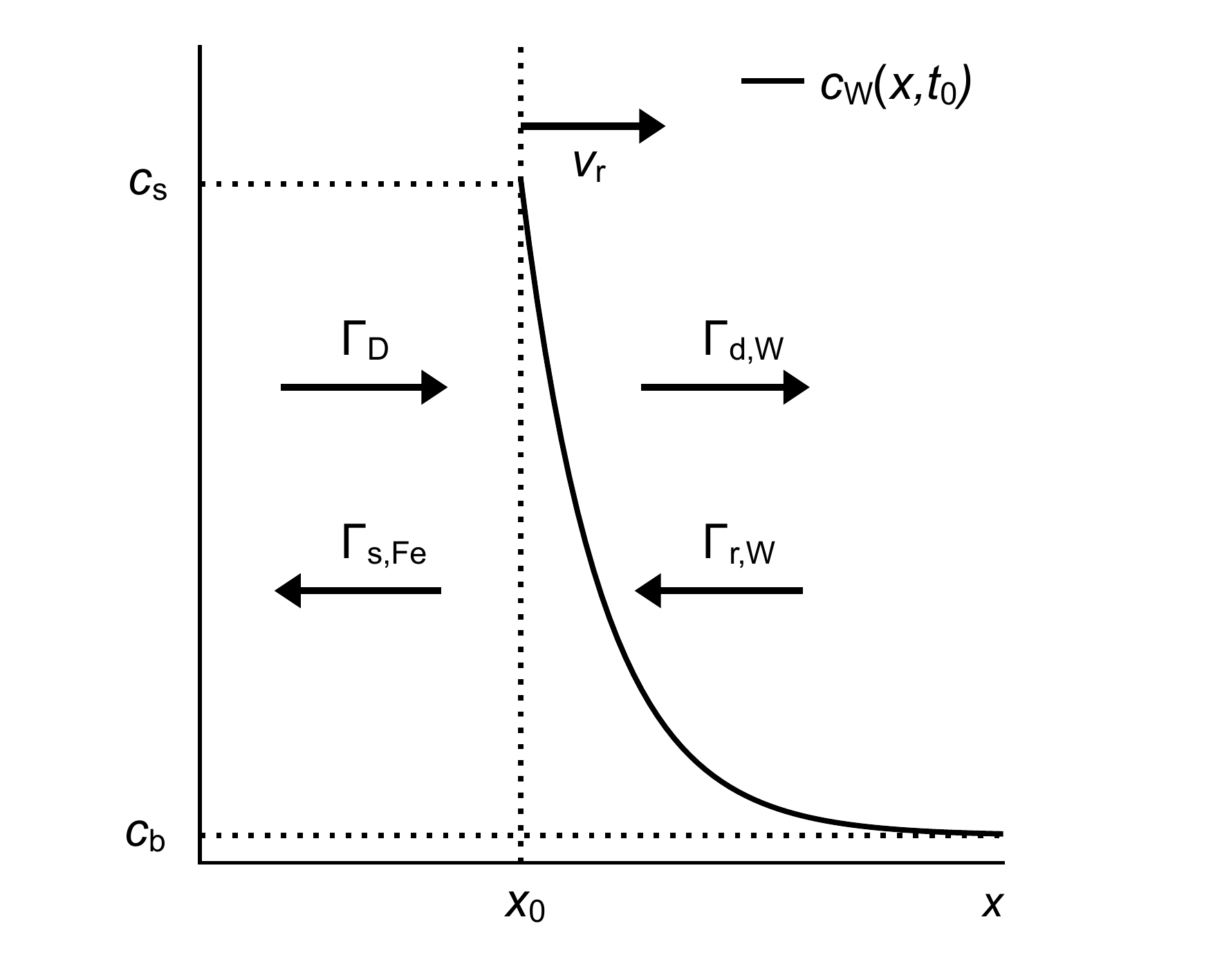}}
\caption{\label{fig-model} \doublespacing
Simple model outlining the various particle fluxes leading to W surface enrichment which have to be accounted for.}
\end{figure}

In the following the diffusion kinetics of W is determined from the LEIS measurements of the temperature-dependent W surface concentration.
The analysis presented here follows general ideas outlined in \cite{pick76,ho78,coll78,swar81}.
For that purpose a simplified model taking into account only the most relevant contributions and restricting the components of the sample to be only Fe and W, is constructed.
The model is sketched in figure \ref{fig-model}.
The exponentially decaying curve is the W concentration $c_{\rm W}(x,t_{0})$ as function of depth $x$ at time $t_{0}$, where $c_{\rm s}$ and $c_{\rm b}$ denote surface and bulk values of W concentration, respectively.
The vertical dotted line indicates the position of the surface of the sample at time $t_{0}$.
$\Gamma_{\rm D}$ is the sputtering D flux, $\Gamma_{\rm s,Fe}$ denotes the flux of sputtered Fe atoms which leave the sample.
Due to continuous sputtering and removal of Fe atoms the surface recesses at a speed $v_{\rm r}$ and W atoms from the bulk enter the surface.
This particular feature is accounted for by introducing the recession flux $\Gamma_{\rm r,W}$.
The last contribution is the temperature driven diffusion of W atoms away from the surface back into the bulk, $\Gamma_{\rm d,W}$.

\begin{figure}[htb]
\resizebox{\figurewidth}{!}{\includegraphics{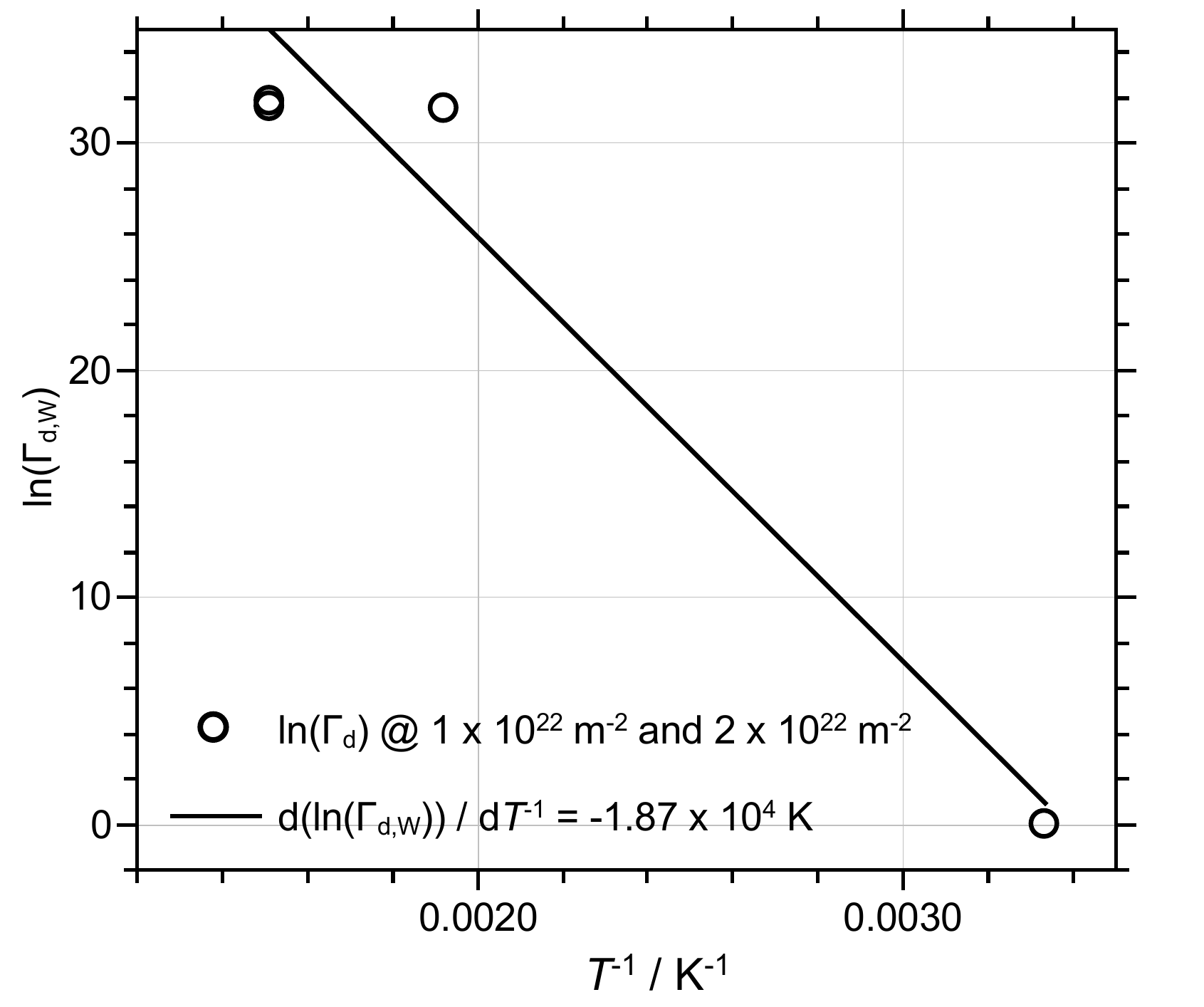}}
\caption{\label{fig-arrhenius} \doublespacing
Arrhenius plot of W diffusion coefficient vs inverse temperature.}
\end{figure}

The above mentioned fluxes can be expressed as follows:
\begin{equation}
\Gamma_{\rm s,Fe} = \Gamma_{\rm D} Y_{\rm Fe} (1 - c_{\rm W})
\end{equation}
where $Y_{\rm Fe}$ is the sputtering yield, i.e. the number of sputtered Fe atoms per incident D atom.
The recession flux of W atoms is given by
\begin{equation}
\Gamma_{\rm r,W} = n c_{\rm W} v_{\rm r} = \Gamma_{\rm D} Y_{\rm Fe} c_{\rm W} (1 - c_{\rm W})
\label{eq-rW}
\end{equation}
where $v_{\rm r}$ is the recession velocity given by $\Gamma_{\rm s,Fe} / n$ with $n$ being the number density of atoms on the surface of the alloy \cite{ho78,coll78}.
The diffusion flux of W from the enriched surface layer into the bulk is
\begin{equation}
\Gamma_{\rm d,W} = -n D \frac{\partial c_{\rm W}}{\partial x}
\label{eq-dW}
\end{equation}
where $D$ denotes the diffusion coefficient.
The difference of W fluxes given by equations \ref{eq-rW} and \ref{eq-dW} provides the effective flux of W atoms to the surface.
Note that the diffusion flux is always smaller than the recession flux.
\begin{equation}
\Gamma_{\rm W} = \Gamma_{\rm r,W} - \Gamma_{\rm d,W} = \frac{{\rm d}\varrho_{\rm W}}{{\rm d}t}
\label{eq-W}
\end{equation}
where $\varrho_{\rm W} = n^{2/3} c_{W}$ is the surface density of W atoms.
The temporal change of the W surface density is given by
\begin{equation}
\frac{{\rm d}\varrho}{{\rm d}t} = n^{2/3} \frac{{\rm d}c_{W}}{{\rm d}t} = n^{2/3} \frac{{\rm d}c_{W}}{{\rm d}\mathit{\Phi}} \frac{{\rm d}\mathit{\Phi}}{{\rm d}t}
\end{equation}
where $\mathit{\Phi}$ is the fluence of D atoms bombarding the surface.
The quantity ${\rm d}c_{W} / {\rm d}\mathit{\Phi}$ is obtained from the measurements shown in figure \ref{fig-enrich} by fitting the data for a specific temperature using the analytical expression $c_{W} = A (1 - B \exp(-C\mathit{\Phi}))$, from which the required derivative is obtained in a straightforward way.
There is another obstacle which cannot be easily worked out.
The spatial gradient of the W concentration is not accessible from the LEIS measurement.
Because equation \ref{eq-W} gives the diffusion flux of W atoms as difference between the recession flux and the measured effective W flux, setting the diffusion flux at room temperature to zero yields the recession flux and permits to determine the diffusion flux at higher temperatures.
Figure \ref{fig-arrhenius} shows the Arrhenius plot of the logarithm of the diffusion flux versus reciprocal temperature using the data for fluences of \SI{1e22}{m^{-2}} and \SI{2e22}{m^{-2}}.
The diffusion activation energy is derived from the slope and equals $E_{\rm a} = \SI{1.6}{eV}$.
Since the diffusion at room temperature is arbitrarily set to zero and may be larger, this value represents an upper limit.

No other data on the W diffusion activation energy in EUROFER could be found in the literature, but there is some data for the activation energy for W diffusion in Fe available.
All of the reported experiments are done at considerably higher temperatures (\SI{>1000}{K}) and analysed either by electron probe micro analysis (EPMA) or by measuring radioactive tracer isotope diffusion.
In these studies activation energies of \SI{2.47}{eV} \cite{albe74}, \SI{2.97}{eV} \cite{take07}, and \SI{2.61}{eV} \cite{pere11} are obtained.
Compared to the present results, these are significantly larger values.
There are various factors which might contribute to the different behaviour.
On the one hand, EUROFER is a more complex alloy with multiple constituents.
On the other hand, the microstructure may play a decisive role.
The EUROFER samples have small grains in the \SI{}{\upmu m} range as is shown in figure \ref{fig-sem}.
The Fe samples investigated in the referred diffusion studies are carefully annealed and have a different structure with larger grains.
An enhanced grain boundary diffusion in EUROFER could be responsible for the observed much lower diffusion activation energy

\newpage
\section{Summary and Conclusion}

The W surface enrichment by preferential sputtering of EUROFER samples has been measured in-situ by LEIS in an ion beam experiment.
Sputtering with \SI{250}{eV} D atoms clearly leads to W enrichment on the surface which is at room temperature not yet saturated at fluences up to \SI{1e23}{m^{-2}}.
Already at a moderately increased temperature of \SI{520}{K} the measured enrichment is only half as large, and at a temperature of \SI{800}{K} no increase in the W surface concentration at the applied sputtering atom flux of \SI{2e18}{m^{-2}s^{-1}} is detectable.
The W surface concentration data measured at various temperatures yield an upper limit for the diffusion activation energy of $E_{\rm a} = \SI{1.6}{eV}$.
This value for EUROFER is considerably smaller than activation energies for W diffusion in Fe (in the range \SI{2.5}{eV} to \SI{3}{eV}) which are reported in the literature.
One important reason for the different finding in the EUROFER system may be the difference in grain structure and size, which shows rather small grains in the \SI{}{\upmu m} range.
An absolute calibration of the LEIS data is done by comparing scattering spectra for pure Fe and pure W.
The surface enrichment has been additionally determined by TOF-MEIS, XPS, and EDX.
All of these methods show smaller values which confirms that the thickness of the enriched surface layer is rather small with a half width in the range of a few \SI{}{nm}.

The results of the present study raises a question for the intended application of EUROFER as a first wall material in a fusion power plant.
Because of the envisaged high operational temperature the strong W diffusion may counteract the beneficial W enrichment which reduces the erosion rate and prolongs the lifetime of the respective wall components.
The present data has been obtained in an ion beam experiment using rather low values of incident particle fluxes compared to the conditions in a fusion reactor.
According to the estimates in \cite{behr03} in a fusion reactor the lower bound for eroding particle fluxes will be about one to two orders of magnitude larger than used in the present laboratory study.
Therefore, the experimental results do not rule out that there still might be W enrichment at higher incident particle fluxes which have to produce W recession fluxes due to preferential sputtering of Fe which exceed the reduction of W surface concentration by diffusion.
Recently, similar experiments showed W surface enrichment after D sputtering of EUROFER samples up to a temperature of \SI{1050}{K} when a four times higher D$^{+}$ ion flux was used \cite{stro18}.
However, the W enrichment will come at the price of larger net wall erosion.
In recessed areas with very low impinging particle fluxes the achievable W surface enrichment may still be influenced by diffusion, but will strongly depend on the operating temperature.

\newpage{}
\section*{Acknowledgements}

We thank Mr. Albert Hiller for the technical support in re-building the apparatus and during the experiments, Mr. Robert Habrichs for setting up the S7 control system of the apparatus, Mr. Stefan Kirtz for installing the personal computers used for data acquisition, Mrs. Beatrix G\"oths for polishing the samples, and Dr. Anne Houben for her assistance with the magnetron sputter coating of the calibration sample.

The operation of the ToF-MEIS setup at Uppsala University has been supported by infrastructure grants from the Swedish Foundation for Strategic Research (SSF) and the Swedish Research Council (VR-RFI). We are thankful to Dr. Daniel Primetzhofer for providing help with the ToF-MEIS measurement and its interpretation.

This work has been carried out within the framework of the EUROfusion Consortium and has received funding from the Euratom research and training programme 2014-2018 under grant agreement No 633053.
The views and opinions expressed herein do not necessarily reflect those of the European Commission.


\newpage
\begin{table}[h]
 \begin{tabular}[h]{ l r r r r r }
  \hline{}
  & at\% Fe & at\% Cr & at\% W & at\% C & at\% O \\
  \hline{}
  all            & 22.6 & 2.0 & 2.5 & 16.4 & 56.5 \\
  neglect O      & 52.0 & 4.6 & 5.8 & 37.6 &      \\
  neglect O \& C & 83.4 & 7.3 & 9.3 &      &      \\
  \hline
 \end{tabular}
 \caption{\label{tab-xps} \doublespacing{}
 Surface concentrations of the EUROFER sample after D atom irradiation with a fluence of $\mathit{\Phi} = \SI{1.1e23}{m^{-2}}$ at room temperature determined with XPS.}
\end{table}


\newpage

\newpage

\newpage

\newpage

\newpage

\newpage

\newpage

\newpage

\newpage

\newpage


\newpage
\begin{thebibliography}{99}
\bibitem{tosc01} R. Toschi et al., \textit{Fusion Eng. Des.}, \textbf{56-57} (2001)  163-172 doi:10.1016/S0920-3796(01)00577-4
\bibitem{bolt02} H. Bolt et al., \textit{J. Nucl. Mater.} \textbf{307-311} (2002) 43-52 doi:10.1016/S0022-3115(02)01175-3
\bibitem{nora03} P. Norajitra et al., \textit{Fusion Eng. Des.} \textbf{69} (2003) 669-673 doi:10.1016/S0022-3115(02)01175-3
\bibitem{lind05} R. Lindau et al., \textit{Fusion Eng. Des.} \textbf{75-79} (2005) 989-996 doi:10.1016/j.fusengdes.2005.06.186
\bibitem{behr03} R. Behrisch et al., \textit{J. Nucl. Mater.} \textbf{313–316} (2003) 388–392 doi:10.1016/S0022-3115(02)01580-5
\bibitem{sugi16} K. Sugiyama et al., \textit{Nucl. Mater. Energy} \textbf{8} (2016) 1-7 doi:10.1016/j.nme.2015.08.001
\bibitem{roth14} J. Roth et al., \textit{J. Nucl. Mater.} \textbf{454} (2014) 1-6 doi:10.1016/j.jnucmat.2014.07.042
\bibitem{goeb84} D. M. Goebel et al., \textit{J. Nucl. Mater.} \textbf{121} (1984) 277-282 doi:10.1016/0022-3115(84)90135-1
\bibitem{moel88} W. M\"oller, \textit{Comp. Phys. Comm.} \textbf{51} (1988) 355-368 doi:10.1016/0010-4655(88)90148-8
\bibitem{kret15} A. Kreter et al., \textit{Fusion Sci. Technol.} \textbf{68} (2015)	8-14 doi:10.13182/FST14-906
\bibitem{rasi17} M. Rasi\'nski et al., \textit{Phys. Scr.} \textbf{T170} (2017) 014036 doi:10.1088/1402-4896/aa8de5
\bibitem{sugi15} K. Sugiyama et al., \textit{J. Nucl. Mater.} \textbf{463} (2015) 272-275 doi:10.1016/j.jnucmat.2014.11.044
\bibitem{stef15} J. Steffens, \textit{Investigations of Preferential Sputtering of Fe/W Samples in the Linear Plasma Device PSI-2 using Glow Discharge Optical Emission Spectroscopy} Master thesis, HHU D\"usseldorf (2015)
\bibitem{rasi16} M. Rasi\'nski et al., \textit{Phys. Scr.} \textbf{T167} (2016) 014013 doi:10.1088/0031-8949/2016/T167/014013
\bibitem{alim16} V. Kh. Alimov et al., \textit{Nucl. Mater. Energy} \textbf{7} (2016) 25-32 doi:10.1016/j.nme.2016.01.001
\bibitem{stro16} P. Str\"om et al., \textit{Nucl. Instr. Meth. B} \textbf{371} (2016) 355-359 doi:10.1016/j.nimb.2015.09.024
\bibitem{stro17} P. Str\"om et al., \textit{Nucl. Mater. Energy} \textbf{12} (2017)	472-477 doi:10.1016/j.nme.2017.03.002
\bibitem{stro18} P. Str\"om et al., \textit{J. Nucl. Mater.} \textbf{508} (2018) 139-146 doi:10.1016/j.jnucmat.2018.05.031
\bibitem{tous16} U. von Toussaint et al., \textit{Phys. Scr.} \textbf{T167} (2016) 014023 doi:10.1016/j.nimb.2015.09.024
\bibitem{nieh93} H. Niehus et al., \textit{Surf. Sci. Rep.} \textbf{17} (1993) 213-303 doi:10.1016/0167-5729(93)90024-J
\bibitem{bron07} H. H. Brongersma et al., \textit{Surf. Sci. Rep.} \textbf{62} (2007) 63-109 doi:10.1016/j.surfrep.2006.12.002
\bibitem{prim11} D. Primetzhofer et al., \textit{Surf. Sci} \textbf{605} (2011) 1913-1917 doi:10.1016/j.susc.2011.07.006
\bibitem{tagl85} E. Taglauer, \textit{Appl. Phys. A} \textbf{38} (1985) 161-170 doi:10.1007/BF00616493
\bibitem{patt67} W. L. Patterson and G. A. Shirn, \textit{J. Vac. Sci. Technol.} \textbf{4} (1967) 343-346 doi:10.1116/1.1492560
\bibitem{yama96} Y. Yamamura and H. Tawara, \textit{At. Data Nucl. Data Tables} \textbf{62} (1996) 149-253 doi:10.1006/adnd.1996.0005
\bibitem{rapp65} D. Rapp et al., \textit{J. Chem. Phys.} \textbf{42} (1965) 4081-4085 doi:10.1063/1.1695897
\bibitem{ruba99} A. V. Ruban et al., \textit{Phys. Rev. B} \textbf{59} (1999) 15990-16000 doi:10.1103/PhysRevB.59.15990
\bibitem{thom86} T. M. Thomas et al., \textit{Surf. Sci.} \textbf{175} (1986) L737-L746 doi:10.1016/0039-6028(86)90225-6
\bibitem{drax03} M. Draxler et al. \textit{Nucl. Instr. Meth. B} \textbf{203} (2003) 218-224 doi:10.1016/S0168-583X(02)02220-6
\bibitem{goeb15} D. Goebl at al., \textit{Nucl. Instr. Meth. B} \textbf{354} (2015) 3-8 doi:10.1016/j.nimb.2014.11.030
\bibitem{kuer13} Ph. K\"urnsteiner et al., \textit{Surf. Sci.} \textbf{609} (2013) 167-171 doi:10.1016/j.susc.2012.12.003
\bibitem{nieh75} H. Niehus et al., \textit{Surf. Sci.} \textbf{47} (1975) 222-233 doi:10.1016/0039-6028(75)90289-7
\bibitem{pouc87} J. L. Pouchou and F. Pichoir \textit{Proceedings of ICXOM} \textbf{11} (1987) 249-253
\bibitem{linn12} M. K. Linnarsson et al., \textit{Rev. Sci. Instrum.} \textbf{83} (2012) 095107 doi:10.1063/1.4750195
\bibitem{bier91} J. P. Biersack et al., \textit{Nucl. Instr. Meth.} \textbf{B61} (1991) 77-82 doi:10.1016/0168-583X(91)95564-T
\bibitem{prim08} D. Primetzhofer et al., \textit{Phys. Rev. Lett.} \textbf{100} (2008) 213201 doi:10.1103/PhysRevLett.100.213201
\bibitem{pick76} H. W. Pickering, \textit{J. Vac. Sci. Technol.} \textbf{13} (1976) 618-621 doi:10.1116/1.569045
\bibitem{ho78}   P. S. Ho, \textit{Surf. Sci.} \textbf{72} (1978) 253-263 doi:10.1016/0039-6028(78)90294-7
\bibitem{coll78} R. Collins, \textit{Radiat. Eff.} \textbf{37} (1978) 13-19 doi:10.1080/00337577808242082
\bibitem{swar81} D. G. Swartzfager et al., \textit{J. Vac. Sci. Technol.} \textbf{19} (1981) 185-191 doi:10.1116/1.571102
\bibitem{albe74} P. Alberry and C. Haworth \textit{Met. Sci.} \textbf{8} 1974 407-412 doi:10.1179/msc.1974.8.1.407
\bibitem{take07} S. Takemoto et al., \textit{Phil. Mag.} \textbf{87} (2007) 1619-1629 doi:10.1080/14786430600732093
\bibitem{pere11} R. A. Perez and D. N. Torres \textit{Appl. Phys. A} \textbf{104} (2011) 329-333 doi:10.1007/s00339-010-6142-x
\end{thebibliography}
\end{document}